\documentclass[aps,prd,a4paper,preprintnumbers,tightenlines,superscriptaddress,
twocolumn,floatfix,showpacs,final,nofootinbib]{revtex4-1}
\usepackage{graphicx}
\usepackage{longtable}
\usepackage{amsmath}
\usepackage{amssymb}
\usepackage{subfigure}
\usepackage{hyperref}
\usepackage{dcolumn}
\usepackage{bm}
\usepackage{color}

\newcommand{\mpl}{{M_{\rm {pl}}}}

\newcommand{\dd}{\, {\rm d}}
\newcommand{\gsim}{\;\mbox{\raisebox{-0.5ex}{$\stackrel{>}{\scriptstyle{\sim}}$}
}\;}
\newcommand{\lsim}{\;\mbox{\raisebox{-0.5ex}{$\stackrel{<}{\scriptstyle{\sim}}$}
}\;}

\newcommand{\ppp}{^\prime}
\newcommand*\colvec[3][]{\begin{pmatrix}\ifx\relax#1\relax\else#1\\\fi#2\\#3\end{pmatrix}}

\newcommand{\nm}{{\mu\nu}}

\newcommand{\ccc}{{_{\rm c}}}
\newcommand{\sss}{{_{\rm s}}}

\newcommand{\GN}{G_{\rm N}}
\newcommand{\GT}{\mathrm{G}^3}
\def\eea{\end{eqnarray}}
\def\bea{\begin{eqnarray}}
\allowdisplaybreaks
\usepackage{enumitem}
\setenumerate[1]{label=(\arabic*)}

\begin{document}
\title{Astrophysical Probes of the Vainshtein Mechanism: Stars and Galaxies}
\author{Kazuya Koyama}
\email[Email:]{kazuya.koyama@port.ac.uk}
\author{Jeremy Sakstein}
\email[Email:]{jeremy.sakstein@port.ac.uk}
\affiliation{Institute of Cosmology and Gravitation,
University of Portsmouth, Portsmouth PO1 3FX, UK}

\begin{abstract}
Ghost-free theories beyond the Horndeski class exhibit a partial breaking of the Vainshtein mechanism inside non-relativistic sources of finite 
extent. We exploit this breaking to identify new and novel astrophysical probes of these theories. Non-relativistic objects feel a gravitational 
force that is weaker than that predicted by general relativity. The new equation of hydrostatic equilibrium is derived and solved to predict 
the modified behaviour of stars. It is found that main-sequence stars are dimmer and cooler than their general relativity counterparts but the red 
giant phase is largely indistinguishable. The rotation curves and lensing potential of Milky Way-like galaxies are calculated. The circular velocities 
are smaller than predicted by general relativity at fixed radius and the lensing mass is smaller than the dynamical mass. We discuss potential 
astrophysical probes of these theories and identify strong lensing as a particularly promising candidate.
\end{abstract}
\maketitle
\section{Introduction}

The acceleration of the cosmic expansion \cite{Riess:1998cb,Perlmutter:1998np} is one of the biggest mysteries in modern physics. For more than 
a decade, there has been a major theoretical and observational effort (see \cite{Copeland:2006wr} for a review) to understand the underlying 
mechanism. Infra-red modifications of gravity are a promising and active area of research (see \cite{Joyce:2014kja} for a review) but any viable 
modification must be compatible with solar system tests. This constraint usually renders any modification irrelevant on cosmological scales due to 
the 
need to tune any new effects to be sub-dominant to general relativity (GR) by five orders-of-magnitude or more. For this reason, theories that 
contain screening mechanisms (see \cite{Jain:2010ka} for a review) such as chameleon gravity \cite{Khoury:2003aq,Khoury:2003rn}, symmetrons 
\cite{Khoury:2010xi} and the Vainshtein mechanism \cite{Vainshtein:1972sx} are particularly promising due to their ability to give rise to order-one 
effects on cosmological scales whilst using non-linear effects to recover GR in the solar system. The first two mechanisms (and their 
generalisations) 
predict a plethora of astrophysical signatures that have been used to place incredibly strong constraints on the model parameters 
\cite{Davis:2011qf,Jain:2011ji,Jain:2012tn,Llinares:2012ds,Vikram:2013uba,Brax:2013uh,Sakstein:2013pda,Terukina:2013eqa,Vikram:2014uza,
Sakstein:2014nfa,Sakstein:2015oqa}. So strong in fact that modified gravity effects cannot drive the cosmic acceleration \cite{Wang:2012kj}. This 
paper is concerned with the Vainshtein mechanism.

Unlike chameleon-like theories, the Vainshtein mechanism is so efficient at screening that, to date, there is an absence of novel small 
scale probes\footnote{With one exception. See \cite{Hui:2012jb}.}. The Vainshtein mechanism features in many modified theories of gravity, the 
quintessential paradigm being galileons \cite{Nicolis:2008in}. Unlike chameleon-like theories, these are defined in the decoupling limit and are the 
most general theory of a scalar field possessing the shift symmetry $\phi\rightarrow\phi+c_\mu x^\mu+b$ in this limit. This symmetry leads to 
higher-order derivative terms in the action such as $(\partial_\mu\phi)^2\Box\phi$ but ensures that the equations of motion are second-order, and 
hence 
the theory is ghost free. In order to examine the cosmology of these theories, one requires a covariantisation so that the theory is defined on 
any background. This covariantisation is not unique and there are many inequivalent theories that give rise to the same theory in the decoupling 
limit. One covariantisation (if only the cubic galileon is present) is DGP gravity \cite{Dvali:2000hr} where the galileon appears in the 
four-dimensional effective theory as the position of the 5D brane. This scenario is ruled out by cosmological observations \cite{Fang:2008kc} and, 
furthermore, the self-accelerating branch has a ghost \cite{Koyama:2007zz}. Another valid covariantisation is ghost free massive gravity 
\cite{deRham:2010kj} and massive bi-gravity \cite{Hassan:2011zd} (see \cite{deRham:2014zqa} for a review of massive gravity). In its simplest form, 
massive gravity does not admit exact Friedmann-Robertson-Walker (FRW) solutions \cite{D'Amico:2011jj} and more complicated forms do not have a 
standard cosmology \cite{Solomon:2014iwa}. Massive bi-gravity does admit FRW solutions but these are unstable at the level of linear perturbations 
\cite{Comelli:2012db,Konnig:2014xva,Lagos:2014lca,Comelli:2015pua}. One may instead try to covariantise the theory directly by promoting the partial 
derivatives to covariant ones. This process spoils the 
ghost-free nature of the theory since couplings to curvature tensors that are zero when the space-time is Minkowski give rise to higher-order 
derivatives of the scalar and the metric. One can remove these from the equations of motion by introducing non-trivial couplings of the scalar to 
curvature tensors in the action. One is then led to Covariant galileons \cite{Deffayet:2009wt}. Unfortunately, this covariantisation is ruled out by 
cosmological probes \cite{Barreira:2012kk,Barreira:2013eea,Barreira:2013jma,Barreira:2013xea}. Covariant galileons are a subset of the Horndeski 
Lagrangian \cite{Horndeski:1974wa}, the most general Lagrangian with a scalar coupled to gravity that results in second-order equations of motion. 
Until recently, this was believed to be the most general ghost-free action for this particle content but \cite{Gleyzes:2014dya,Gleyzes:2014qga} (see 
also \cite{Zumalacarregui:2013pma}) have shown that there are healthy extensions of this whereby the equations of motion are at first glance 
higher-order but reduce to second-order after the application of hidden constraints. In particular, the n\"{a}ive covariantisation of galileons 
discussed above falls into this class of models. 

Interestingly, \cite{Kobayashi:2014ida} have shown that the Vainshtein mechanism is only partially operational in theories beyond Horndeski. Whereas 
it operates perfectly outside sources there is an incomplete suppression of the fifth-force in the interior of extended sources due to the 
time-dependence of the cosmological field (see \cite{Sakstein:2014isa,Sakstein:2014aca} for other examples of how cosmological time-dependence can be 
important on solar system scales in extended theories of gravity). This opens up the possibility of testing these theories on small scales using 
astrophysics. The aim of this paper is to discern the new novel probes that may potentially be used to probe these theories.

We begin by looking at the effects on stellar structure. After finding the new equations for the metric potentials $\Phi$ and $\Psi$ inside 
sources, the equation of hydrostatic equilibrium predicted in these theories is derived. The theory introduces one additional dimensionless 
parameter, $\Upsilon$, that depends on $\dot{\phi}$, the time-derivative of the cosmological scalar and $\Lambda$, the mass-scale appearing in the 
quartic galileon Lagrangian. In order to gain some analytic and physical insight into how the properties of stars are altered in these 
theories, polytropic models (with an equation of state $P\propto\rho^\gamma$) are examined. These are useful because they decouple the effects of 
gravitational physics from those due to baryonic physics. The equations of stellar structure remain homology-invariant, which allows a 
Lane-Emden-like equation governing the structure of polytropic stars to be found. Main-sequence stars are 
well-described by $\gamma=4/3$ polytropes and so we proceed to solve the Lane-Emden equation for this value. It is found that main-sequence stars are 
less luminous than their GR counterparts at fixed mass. In order to provide more accurate predictions that can be compared with data, we have 
modified the stellar evolution code MESA \cite{Paxton:2010ji,Paxton:2013pj} to include the new equation of hydrostatic equilibrium predicted by these 
theories. The code reveals that for low mass stars ($\sim1M_\odot$) the main effect of the modifications is that equivalent stars have higher 
effective temperatures at fixed luminosity. This effect is large on the main-sequence but is only minor on the red giant branch and is degenerate 
with 
keeping GR as the theory of gravity and increasing the metallicity. Higher mass stars exhibit similar features except the luminosity dimming is more 
pronounced.

Finally, we turn our attention to galactic probes. We compute the modified rotation curves and lensing potentials for Milky Way-like galaxies. It is 
found that non-relativistic objects have smaller circular velocities at fixed radii than predicted by GR and that light is lensed less strongly. We 
comment on the use of strong lensing as a potential probe of this effect.

This paper is organized as follows: Quartic galileons and the Vainshtein mechanism are reviewed in the next section where the equations governing the 
metric potentials $\Phi$ and $\Psi$ are derived. In section \ref{sec:ss} the new equation of hydrostatic equilibrium is derived. It is used to find 
the modified Lane-Emden equation governing the structure of polytropic stars and we subsequently specialise to stars with equations of state of the 
form $P\propto \rho^{4/3}$ i.e. main-sequence stars. The resultant equations are solved to show that main-sequence polytropic stars are dimmer than 
their GR counterparts. In the remainder of the section, the stellar structure code MESA is used to produce realistic predictions and it is found that 
in addition to being dimmer, main-sequence stars are cooler (in the sense that they have smaller effective temperatures). Galactic effects are 
studied in section \ref{sec:galtests} where we examine the rotation curves and lensing potentials of Milky Way-like galaxies. The circular velocities 
of non-relativistic objects are smaller than the GR prediction at fixed radius and light is less strongly lensed. We discuss our results and conclude 
in section \ref{sec:concs}. The metric signature is $(-,+,+,+)$ and the Planck mass $\mpl^2=(8\pi\GN)^{-1}$.

\section{Modified Gravity Theories and the Vainshtein Mechanism}\label{sec:Vain}

Theories that screen using the Vainshtein mechanism are abundant in the literature (see \cite{Joyce:2014kja} for example) and galileons are the 
most common paradigm by far. These are defined in the decoupling limit by the following Lagrangian\footnote{We ignore the tadpole term here.}:
\begin{equation}
\mathcal{L}=\mpl^2\sum_i\frac{\mathcal{L}_i}{\Lambda_i^{2(i-2)}}+\alpha\phi T+\frac{T^\nm\partial_\mu\phi\partial_\nu\phi}{\mathcal{M}^4},
\end{equation}
where $T^\nm$ is the energy-momentum tensor in the matter sector and $T$ is its trace.
\begin{align}
\mathcal{L}_2&\equiv  X\\
\mathcal{L}_3&\equiv X\Box\phi-\phi_\mu\phi^{\mu\nu}\phi_\nu\\
\mathcal{L}_4&\equiv\nonumber-X\left[(\Box\phi)^2-\phi_{\mu\nu}\phi^{\mu\nu}\right]\\&-\left(\phi^\mu\phi^\nu\phi_{\mu\nu}\Box\phi-\phi^\mu
\phi_{\mu\nu}\phi_\rho\phi^{\rho\nu}\right),\\ \mathcal{L}_5&\equiv-2X\left[\left(\Box\phi\right)^3-3\phi_\nm\phi^\nm\Box\phi+2\phi_\nm\phi^{\nu\rho}
\phi^\mu_{\,\,\rho}\right]\nonumber\\&-\frac{3}{2}\left(\left(\Box\phi\right)^2\phi^\mu\phi^\nu\phi_\nm\nonumber-2\phi_\mu\phi^\nm\phi_{\nu\rho}
\phi^\rho\right.\\&\left.\,\,\,-\phi_\nm\phi^\nm\phi_\rho\phi^{\rho\sigma}\phi_\sigma+2\phi_\mu\phi^\nm\phi_{\nu\rho}\phi^{\rho\sigma}
\phi_\sigma\right).
\end{align}
Here $\phi_{{\mu}_1\ldots{\mu}_n}=\nabla_{{\mu}_1}\ldots\nabla_{{\mu}_n}\phi$ and $X=-1/2\partial_\mu\phi\partial^\mu\phi$\footnote{Note that our 
definition of $X$ differs from \cite{Gleyzes:2014dya,Gleyzes:2014qga} by a factor of $-2$ but conforms with \cite{Kobayashi:2014ida}. Furthermore, 
our fields are dimensionless, which conforms with neither previous work but has the advantage that the equation of motion for $\phi$ is similar to 
the Poisson equation.}. 
The theory we will study below does not have a conformal coupling 
$\alpha$ but instead sources the galileon through couplings to the curvature tensor present when the quartic term is covariantised. 

Despite being higher-order in the action, the equations of motion are precisely second-order and so there are no ghost 
instabilities. Whilst fine around flat space, the final two Lagrangians do not give rise to second-order equations of motion about an arbitrary 
space-time because their variation leads to couplings of the field to curvature tensors that are identically zero when the space-time is 
Minkowski\footnote{This is due to the relation $\left[\nabla_\mu\phi\nabla_\nu\phi\right]\phi^\alpha=R^\alpha_{\beta\nm}\phi^\beta$ required when 
varying with respect to the field.}. These terms always contain third-order (or higher) derivatives of the metric or fields. In fact, the Lagrangians 
above are not unique and there is an infinite number of equivalent descriptions that are related via integration by parts. Indeed, one can remove the 
final term in $\mathcal{L}_3$ and the terms in the curly brackets in $\mathcal{L}_4$ and $\mathcal{L}_5$. One way to restore the ghost-free nature of 
the theory is to start from these simpler forms and add additional couplings of the fields to curvature tensors so that the higher-order derivatives 
exactly cancel. With this procedure, one is led to covariant galileons \cite{Deffayet:2009wt}. These are a subset of the Horndeski Lagrangian 
\cite{Horndeski:1974wa}, the most general Lagrangian with second-order equations of motion. Recently, \cite{Gleyzes:2014dya,Gleyzes:2014qga} (see 
also 
\cite{Zumalacarregui:2013pma}) have shown that there exist Lagrangians that give rise to higher-order equations that are still ghost-free due to the 
existence of hidden constraints. The n\"{a}ive covariantisation of $\mathcal{L}_4$ and $\mathcal{L}_5$ are a subset of the most general Lagrangian 
and 
so this presents us with a new and interesting theory that screens using the Vainshtein mechanism that has yet to be explored.  In fact, one can show 
that the static, spherically symmetric Vainshtein solution is unstable to perturbations when $c_5\ne0$ \cite{Koyama:2013paa} and so from here on we 
will investigate the Lagrangian 
\begin{equation}\label{eq:L4}
\frac{\mathcal{L}}{\sqrt{-g}}=\mpl^2\left[\frac{R}{2}+X+\frac{\mathcal{L}_4}{\Lambda^4}\right].
\end{equation}
\cite{Gleyzes:2014qga} have dubbed the entire class of beyond Horndeski theories $\GT$ and so we will refer to this theory as the 
$\GT$-galileon.

Before examining the breaking of the Vainshtein mechanism in this covariantisation, we briefly review the screening of the quartic galileon in the 
decoupling limit for comparative purposes. The equation of motion for static, spherically symmetric 
over-densities, $T^{\nm}=\mathrm{diag}(\rho(r),0,0,0)$, is
\begin{equation}\label{eq:vss}
\frac{1}{r^2}\frac{\dd}{\dd r}\left[r^2\phi\ppp+\frac{2}{\Lambda^4} {\phi\ppp}^3\right]=8\alpha\pi G\rho.
\end{equation}
The first term is simply $\nabla^2\phi$ and is due to the quadratic term in the Lagrangian, whilst the second term is due to the quartic Lagrangian. 
The fifth-force $F_5$ due to the conformal coupling (see \cite{Sakstein:2014isa,Sakstein:2014aca} for a discussion of the disformal coupling) is 
$F_5=-\alpha \phi\ppp$ and so one can integrate (\ref{eq:vss}) to find the algebraic relation
\begin{equation}\label{eq:vss2}
 F_5 + \frac{2}{\alpha^2\Lambda^4r^2}F_5^3 = 2 \alpha^2 F_{\rm N}, 
\end{equation}
where $F_{\rm N}$ is the Newtonian force. One can see that on large enough length scales (to be made precise shortly) $F_5=2\alpha^2F_{\rm N}$ is a 
consistent solution of the equation and the force is unscreened. On smaller length scales, one has 
\begin{equation}
 \frac{F_5}{F_N}=\left(\frac{r}{r_{\rm V}}\right)^2\quad r_{\rm V}^3\equiv 
2\alpha\frac{GM}{\Lambda^2},
\end{equation}
where we have assumed a point mass source $\rho(r)=M\delta^{(3)}(r)$ in order to provide a concrete expression. The Vainshtein radius $r_{\rm V}$ 
defines the transition between the screened and unscreened regime. When $r\gg r_{\rm V}$, the force is unscreened and when the converse is true one 
has $F_{5}\ll F_{\rm N}$ and the force is screened. Typically, an object characterised by some length scale $R$ has a Vainshtein radius $r_{\rm V}\gg 
R$, for example, the Vainshtein radius of the Sun encompasses the entire solar system \cite{Afshordi:2008rd,Andrews:2013qva}. For this reason, the 
Vainshtein mechanism is very efficient at screening and there have been few astrophysical signatures to date (see 
\cite{Hui:2012jb,Hiramatsu:2012xj,Falck:2014jwa} 
for some exceptions to this). Below, we will examine the non-relativistic limit of the $\GT$-galileon and show that the Vainshtein mechanism is only 
partially effective.

\subsection{The Non-Relativistic Limit}
We wish to derive the equations governing Newtonian perturbations about an FRW space-time sourced by some non-relativistic density profile $\rho(r)$. 
To this end, we expand the field about its cosmological value $\phi(r,t)=\phi_0(t)+\pi(r,t)$ and the metric as 
\begin{equation}\label{eq:cngauge}
 \dd s^2=-\left[1+2\Phi(r,t)\right]\dd t^2 +a^2(t)\left[1-2\Psi(r,t)\right]\delta_{ij}\dd x^i\dd x^j.
\end{equation}
The radial coordinate is 
$r=a\sqrt{\delta_{ij}x^ix^j}$. In what follows, we will assume that the local effects are due primarily to the cosmological 
boundary condition on $\phi$ and not due to effects coming from the FRW metric. To this end we will ignore all factors of $a(t)$ and 
$H(t)$\footnote{This is an ad hoc assumption that must be tested for self-consistency. In fact, if one were to include these factors, the only change 
would be in how the final new parameter $\Upsilon$ is defined in terms of the parameters appearing in the fundamental theory. Since we will always 
work with $\Upsilon$ and not $\dot{\phi}_0$ or $\Lambda$ we are not missing any potential new effects by making this assumption.}. We also ignore any 
time-dependence since we are interested in static situations in this work. Next, we must solve the equations of motion for the dynamics of $\Phi$, 
$\Psi$ and $\pi$. 

This has already been studied by \cite{Kobayashi:2014ida} for the full set of beyond Horndeski theories. There, they expanded the action to the 
relevant order in metric and field perturbations in order to derive an effective non-relativistic action. One must be careful when gauge-fixing the 
action. There are certain criteria for doing so in order that information is not lost in the resulting equations of motion and, as discussed by 
\cite{Lagos:2013aua}, fixing the conformal Newtonian gauge (\ref{eq:cngauge}) at the level of the action may result in equations of motion that are 
not equivalent to those found from the gauge fixed forms of the equations found using the full action. For this reason, we have derived the 
non-relativistic limit by gauge-fixing the equations of motion and not the action. This derivation can be found in Appendix \ref{sec:eoms}. In fact, 
our equations are identical to those of \cite{Kobayashi:2014ida} when their parameters are chosen to match ours but there is no underlying reason for 
this and, indeed, it is not necessarily the case that this is the case for more general models.

In what follows, we will use the variables (which are similar to those used by \cite{Kobayashi:2014ida})
\bea
& &
x\equiv \frac{\pi^\prime}{r}\,,~~
y \equiv \frac{\Phi\ppp}{r}\,, 
\nonumber \\
& &
z \equiv \frac{\Psi\ppp}{r}\,,~~
A\equiv\frac{M(r)}{8\pi\mpl^2 r^3}\,,\label{eq:defs}
\eea
where a prime denotes a derivative with respect to the 
radial coordinate and $M(r)$ is the mass enclosed within a sphere of radius $r$. Using these variables, one finds the following equations in the 
non-relativistic limit (after integrating the equations of motion by parts, see Appendix \ref{sec:eoms}):
\begin{align}
 z+\frac{5\varepsilon}{2}x^2&=A,\label{eq:z}\\
 y-z-\frac{\varepsilon}{2}x^2-\varepsilon x\left(rx\ppp+x\right)&=0\quad\textrm{and}\label{eq:y}\\
 x-4\frac{x^3}{\Lambda^4}+10\varepsilon xy+2\varepsilon x\left(2z+rz\ppp\right)&=0,\label{eq:x}
\end{align}
where
\begin{equation}
 \varepsilon\equiv\frac{\dot{\phi}_0^2}{\Lambda^4}.
\end{equation}
One can then use equations (\ref{eq:z}) and (\ref{eq:y}) in equation (\ref{eq:x}) to find an algebraic equation for $x$:
\begin{equation}
 \left(\frac{4}{\Lambda^4}+20\varepsilon^2\right)x^2-2\varepsilon\left[\left(7A+rA\ppp\right)\right] - 1=0\label{eq:X}.
\end{equation}
Note that $x\ppp$ does not appear in this equation; this is the \textit{hidden constraint} of the beyond Horndeski action. Despite the fact that the 
equations of motion appear third-order, only two integrations are required to fully solve the system; the hidden constraint has allowed us to find a 
suitable linear combination of the equations that are second-order. Taking $A\gg1$\footnote{We 
assume $rA\ppp\sim A$.} one finds\footnote{The careful reader will notice that (\ref{eq:X}) is in fact a cubic equation before cancelling a common 
factor of $x$ and that $x=0$ appears to be a consistent solution. In fact, there is a small constant term in the cubic equation that is Hubble 
suppressed \cite{Kobayashi:2014ida} and so this solution cannot be realised in practice. Furthermore, we are using the $\GT$-galileon as an example 
of the breaking of the Vainshtein mechanism in beyond Horndeski theories. The equations we will ultimately obtain are correct for any beyond 
Horndeski theory and are obtained by taking the $x\ne0$ branch of solutions. See \cite{Kobayashi:2014ida} for the general derivation.}
\begin{equation}\label{eq:xsol}
 x^2=\varepsilon\Lambda^4\frac{7A+rA\ppp}{10\Upsilon+2}.
\end{equation}
At this point we introduce the new parameter
\begin{equation}
 \Upsilon\equiv \varepsilon^2\Lambda^4=\left(\frac{\dot{\phi_0}}{\Lambda}\right)^4.
\end{equation}
This is the parameter that quantifies the size of the new effects presented here and we will work with it exclusively from here on. Using 
(\ref{eq:xsol}), the solutions for $y$ and $z$ are
\begin{align} 
y&=\frac{A}{1+{5}\Upsilon}+\frac{\Upsilon}{4(1+5\Upsilon)}\frac{\left(r^3A\right)^{\prime\prime}}{r}\quad\textrm{and}       
\label{eq:ysol}\\
 z&=\frac{A}{1+5\Upsilon}-\frac{5\Upsilon}{4(1+5\Upsilon)}\frac{\left(r^3A\right)\ppp}{r^2}.\label{eq:zsol}
\end{align}
One would ideally identify Newton's constant as
\begin{equation}\label{eq:GNdef}
 G_{\rm N}\equiv \frac{G}{1+5\Upsilon},
\end{equation}
but this definition is subtle since we do not know the solution for the metric potentials outside of the source. Traditionally, one works in a gauge 
where, to Newtonian order,
\begin{equation}\label{eq:gammadef}
 \dd s^2= -\left(1+\frac{2\GN M(R)}{r}\right)+\left(1-2\gamma\frac{\GN M(R)}{r}\right)
\end{equation}
outside the source, which defines $\GN$. We will show in Appendix \ref{sec:VM} that (\ref{eq:GNdef}) is indeed the correct definition of $\GN$. This 
then allows us to write (\ref{eq:ysol}) and (\ref{eq:zsol}) in the more useful form
\begin{align}
 \frac{\dd\Phi}{\dd r}&=\frac{\GN M(r)}{r^2}+\frac{\Upsilon}{4}\GN M^{\prime\prime}(r)\label{eq:dPhi}\\
 \frac{\dd \Psi}{\dd r}& = \frac{\GN M(r)}{r^2}-\frac{5\Upsilon}{4}\frac{\GN M\ppp(r)}{r}\label{eq:dPsi}. 
\end{align}
These will be our starting point when deriving the new properties of stars and galaxies. Before doing so 
however, it is worth pausing to note three 
important properties of these equations. First, the extra terms that appear are not due to contributions from $x\sim \pi\ppp$; these terms have been 
suppressed by the Vainshtein mechanism. Instead, they are due to the cosmological time-derivatives that the Vainshtein mechanism fails to suppress. 
This behaviour has previously been noted in the case of covariant quartic galileons \cite{Barreira:2013xea}, however there the effects were simply a 
time-dependent rescaling of Newton's constant. Second, we will show in Appendix \ref{sec:VM} that the PPN parameter 
$\gamma=1$ and this means that, a priori, there are currently no constraints on the parameter $\Upsilon$\footnote{It is known that the theories here 
are related to Horndeski theories via a disformal transformation \cite{Gleyzes:2014dya} and so the constraints of \cite{Kaloper:2003yf,Brax:2014vva} 
may apply to $\Upsilon$ once one has accounted for the Vainshtein mechanism and specified a cosmology.}. In practice, 
we expect it to be small and one of the aims of this work is to discern the typical values which lead to large deviations from GR. Third, the final 
term in equation (\ref{eq:dPhi}) is typically negative provided that the density decreases outwards (as it does in most astrophysical systems). Since 
$\dd \Phi/\dd r$ governs the motion of non-relativistic particles, the theory predicts that any such objects moving inside a larger source will feel 
a 
reduced strength of gravity than that predicted by GR.

\section{Stellar Structure Tests}\label{sec:ss}

In this section we derive the new stellar structure equations predicted by the theory and use them to predict novel phenomena that may be used as 
observational probes. We note that, while the derivation of the equation for $\dd\Phi/\dd r$ used the $\GT$-galileon as its starting point, the final 
form applies to any beyond Horndeksi theory \cite{Kobayashi:2014ida}. All that changes is the definition of the parameter $\Upsilon$ in terms of the 
fundamental parameters that appear in the Lagrangian. For this reason, equation \eqref{eq:dPhi}, and the results we obtain here, are completely 
general and one may treat $\Upsilon$ as a free parameter\footnote{This is because non-relativistic objects such as main-sequence stars respond to 
$\Phi$ only. In the most general theory, the parameter appearing in the equation governing $\Phi$ may differ from the one appearing in the equation 
governing $\Psi$, and so one can have two free parameters in general.}.

Since we are working in the Jordan frame, the energy-momentum tensor of matter is covariantly conserved and so one has $\nabla_\mu T^{\mu\nu}=0$. 
Taking the energy-momentum tensor to be that of a static, spherically symmetric non-relativistic source, the Euler equation found from this 
conservation law is
\begin{equation}
 \frac{1}{\rho}\frac{\dd P}{\dd r}=-\frac{\dd \Phi}{\dd r}.
\end{equation}
using equation (\ref{eq:dPhi}), one finds the new hydrostatic equilibrium equation, which governs the structure of stars:
\begin{equation}
 \label{eq:HSE}
 \frac{\dd P}{\dd r}=-\frac{\GN M(r)\rho(r)}{r^2}-\frac{\Upsilon}{4}\GN\rho(r) M^{\prime\prime}(r).
\end{equation}
This is the only modification to the stellar structure equations in this theory. The conservation of mass is unaltered:
\begin{equation}\label{eq:M'}
 \frac{\dd M}{\dd r}=4\pi r^2 \rho(r)
\end{equation}
and for later convenience one can differentiate this to find
\begin{equation}\label{eq:M''}
 \frac{\dd^2 M}{\dd r^2}=8\pi r \rho+4\pi r^2 \frac{\dd \rho}{\dd r}.
\end{equation}
The other two stellar structure equations do not depend on the theory of gravity; they are the radiative transfer and energy generation equations 
respectively \cite{prialnik2000introduction}:
\begin{align}
 \frac{\dd T}{\dd r}&=-\frac{3}{4a}\frac{\kappa}{T^3}\frac{\rho L}{4\pi r^2}\quad\textrm{and}\label{eq:RTE}\\
 \frac{\dd L}{\dd r}&=4\pi r^2\epsilon(r),
\end{align}
where $L$ is the luminosity, $\kappa$ is the opacity, $\epsilon(r)$ is the energy generation rate per unit mass and $a$ is the constant relating the 
radiation pressure to the temperature (see below). Equation (\ref{eq:HSE}) is all that is required to study the structure of non-relativistic stars 
in these theories. We will do this below, first for simple polytropic models and then for realistic stars found using a consistent numerical 
implementation of the new hydrostatic equilibrium equation. The former models are not powerful enough to allow a comparison with real data but they 
are useful in that they decouple the non-gravitational physics, which allows one to gain an analytic handle on the new effects arising in these 
theories without the complications arising from thermodynamic effects. Furthermore, main-sequence stars are well-approximated by certain Lane-Emden 
models and this will allow us to discern the new features one would expect in stars like the Sun. Lane-Emden models have been particularly useful in 
understanding how stars behave in chameleon (and other similar) theories of gravity \cite{Davis:2011qf,Sakstein:2013pda,Sakstein:2014nfa}.

\subsection{Polytropic Models}

In the absence of any knowledge of the micro-physics, the stellar structure equations do not close and one must supply and equation of 
state relating $P$ and $\rho$. Polytropic models are defined by the equation of state
\begin{equation}\label{eq:poly}
P=K \rho^{\frac{n+1}{n}}. 
\end{equation}
$n$ is known as the \textit{polytropic index}. Main-sequence stars are well-described by $n=3$ and fully convective stars are well-modelled by 
$n=1.5$. In GR, polytropic models lead to self-similar equations of stellar structure and since the new parameter $\Upsilon$ is dimensionless the 
same is true in this theory. We can then scale out the dimensionful quantities in the theory by writing $r=r\ccc\xi$ $\rho=\rho\ccc\theta(\xi)^n$ 
and $P=P\ccc\theta(\xi)^{n+1}$, where $P\ccc$ and $\rho\ccc$ are the central pressures and densities and
\begin{equation}\label{eq:rc}
 r\ccc^2\equiv\frac{(n+1)P\ccc}{4\pi \GN \rho\ccc^2}.
\end{equation}
Note that $P\ccc$ and $\rho\ccc$ are related via $P\ccc=K\rho\ccc^{(n+1)/n}$. One can then insert equation (\ref{eq:poly}) into equation 
(\ref{eq:HSE}) and use equations (\ref{eq:M'}) and (\ref{eq:M''}) to find
\begin{equation}\label{eq:MLE}
 \frac{1}{\xi^2}\frac{\dd}{\dd \xi}\left[\left(1+{\frac{n}{4}\Upsilon\xi^2\theta^{n-1}}\right)\xi^2\frac{\dd \theta}{\dd 
\xi}+\frac{\Upsilon}{2}\xi^3\theta^n\right]=-\theta^n.
\end{equation}
When $\Upsilon=0$ this equation describes the structure of polytropic spheres collapsing under gravity described by GR. In this case, it is known 
as the Lane-Emden equation and so we refer to (\ref{eq:MLE}) and the \textit{modified Lane-Emden equation} (MLE). The boundary conditions for the 
dimensionless density $\theta$ are $\theta(0)=1$ (so that $\rho(0)=\rho\ccc$) and $\theta\ppp(0)=0$, which is a requirement of spherical symmetry. 
The radius of the star is found from the value of $\xi_R$ defined by $\theta(\xi_R)=0$ so that $R=r\ccc \xi_R$. One can then see that the extra terms 
in equation (\ref{eq:MLE}) vanish at the surface of the star and for this reason, stars in this theory share many features with GR. In particular, 
the mass of the star is
\begin{align}\label{eq:mass}
 M&\nonumber=\int_0^R\dd r 4\pi r^2\rho(r)\\&=4\pi r\ccc^3\rho\ccc\int_0^{\xi_R}\dd \xi \xi^2 \theta(\xi)^n=4\pi r\ccc^3\rho\ccc\omega_R,
\end{align}
where
\begin{equation}
 \omega_R\equiv - \xi_R^2\left.\frac{\dd \theta}{\dd \xi}\right\vert_{\xi_R}.
\end{equation}
This is precisely the same formula that is found in GR. The only difference being that $\omega_R$ will be different depending on the value of 
$\Upsilon$. In particular, this means that whereas a star with fixed central density will have different masses between the two theories, the 
mass-radius relationship $R\propto M^{(n-1)/(n-3)}$ predicted by GR is unaltered.

One can in principle solve equation (\ref{eq:MLE}) to find the new properties of many polytropes. The goal of this paper is to predict the new 
observational features this theory predicts and so we will now specialise to the case of main-sequence stars.

\subsubsection{Main Sequence Stars}

We will investigate the new properties of these stars using the \textit{Eddington standard model}. The key assumption of this model is that the 
specific entropy gradient ($\propto T^3/\rho$) is constant, which allows us to decouple the radiative transfer and energy generation equations. The 
opacity is taken to be constant and we assume that the pressure supporting the star is due to the thermodynamic properties of the gas,
\begin{equation}\label{eq:Pgas}
 P_{\rm gas}=\frac{\rho k_{\rm B} T}{\mu m_{\rm H}},
\end{equation}
where $\mu$ is the mean molecular mass, and the radiation pressure from photons generated by hydrogen burning in the core,
\begin{equation}\label{eq:Prad}
 P_{\rm rad}=\frac{1}{3}a T^4,
\end{equation}
so that the total pressure is 
\begin{equation}\label{eq:PT}
 P=\frac{\rho k_{\rm B} T}{\mu m_{\rm H}}+\frac{1}{3}a T^4.
\end{equation}
Defining the quantity $\beta\equiv P_{\rm gas}/P$, one can equate $ P_{\rm gas}$ with $P_{\rm rad}/(1-\beta)$ using (\ref{eq:Pgas}) 
and (\ref{eq:Prad}) to find
\begin{equation}\label{eq:beta}
 \frac{1-\beta}{\beta}=\frac{3 k_{\rm B}\rho}{a\mu m_{\rm H} T^3}.
\end{equation}
The assumption that the specific entropy gradient is constant ensures that $\beta$ is a constant, in which case one can use equation (\ref{eq:beta}) 
in equation (\ref{eq:PT}) to find
\begin{equation}\label{eq:polyMS}
 P=K(\beta)\rho^{\frac{4}{3}}
\end{equation}
with 
\begin{equation}\label{eq:Kbeta}
 K(\beta)=\left(\frac{3}{a}\right)^{\frac{1}{3}}\left(\frac{k_{\rm B}}{\mu m_{\rm 
H}}\right)^{\frac{4}{3}}\left(\frac{1-\beta}{\beta^4}\right)^{\frac{1}{3}}.
\end{equation}
Equation (\ref{eq:polyMS}) is the equation of state of an $n=3$ polytrope and so we can describe these stars by solving the modified Lane-Emden 
equation. One important observational property of main-sequence stars is their Luminosity, which we can predict as follows. Differentiating equation 
(\ref{eq:Prad}) and substituting it into the hydrostatic equilibrium equation (setting $P=(1-\beta)P_{\rm rad}$) one finds (at the surface of the 
star)
\begin{equation}\label{eq:lum}
 L=\frac{4\pi \GN M(1-\beta)}{\kappa}.
\end{equation}
This is identical to the expression found in general relativity and so any observational differences (at fixed mass), if any, are encoded in $\beta$. 
One then requires an expression for $\beta$. This can be found as follows: Inserting the definition of $r\ccc$ (\ref{eq:rc}) and equation 
(\ref{eq:Kbeta}) into equation (\ref{eq:mass}) one finds the modified form of Eddington's quartic equation:
\begin{equation}\label{eq:Equart}
 \frac{1-\beta}{\beta^4}=\left(\frac{\bar{\omega}_R}{\omega_R}\right)^2\left(\frac{M}{M_{\rm Edd}}\right)^2,
\end{equation}
where 
\begin{equation}\label{eq:eddmass}
 M_{\rm Edd}\equiv \frac{4\bar{\omega}_R}{\sqrt{\pi}\GN^{\frac{3}{2}}}\left(\frac{k_{\rm B}}{\mu m_{\rm 
H}}\right)^2\left(\frac{3}{a}\right)^{\frac{1}{2}}
\end{equation}
is the \textit{Eddington mass} and $\bar{\omega}_R\approx2.018$ is the value of $\omega_R$ when $\Upsilon=0$ (i.e. the result in GR). Numerically, 
one has $M_{\rm Edd}\approx18.3\mu^{-2}$ and from here on we take $\mu=0.5$, corresponding to fully ionised hydrogen, which is appropriate for main 
sequence stars burning on the PP-chains. Equation (\ref{eq:Equart}) then shows that all of the modified gravity effects are encoded in the new value 
of $\omega_R$, which is a number describing the change in the structure of the star.

\begin{figure}[ht]\centering
\includegraphics[width=0.45\textwidth]{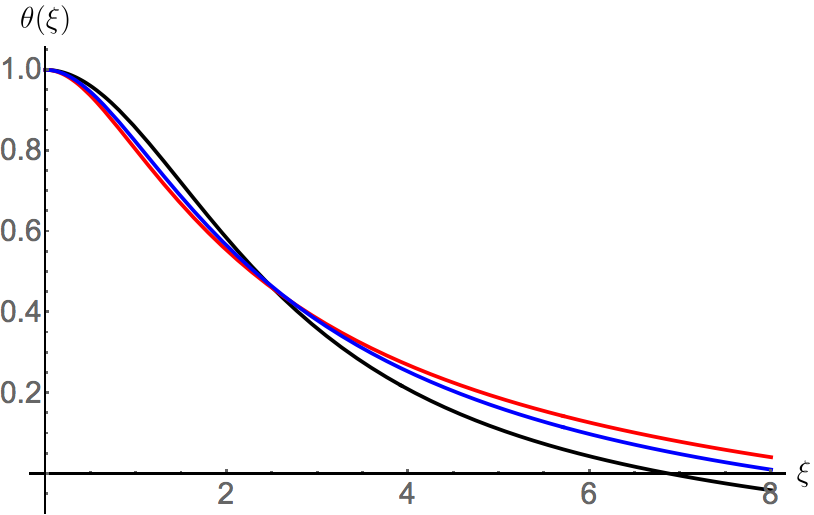}
\caption{The $n=3$ solutions of the Lane-Emden equation (black) and the MLE with $\Upsilon=0.3$ (blue) and 
$\Upsilon=0.5$ (red).}\label{fig:lesolms}
\end{figure}

\begin{figure}[ht]\centering
\includegraphics[width=0.45\textwidth]{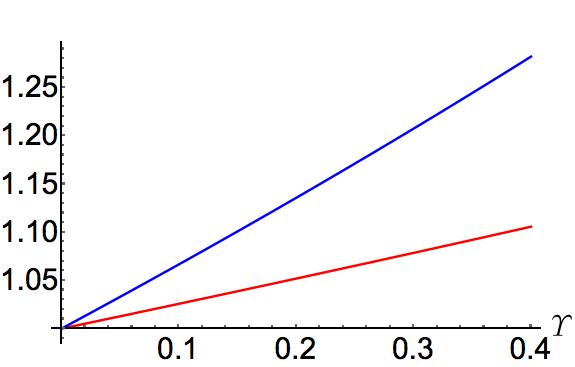}
\caption{$\bar{\omega}_R/\omega_R$ (red) and $\bar{\xi}_R/\xi_R$ (blue) as a function of $\Upsilon$ for $n=3$ solutions of the MLE.}\label{fig:om}
\end{figure}

We are now in a position to integrate the MLE numerically and derive the new properties of main-sequence stars. It is well-known that there are 
subtleties with integrating the Lane-Emden equation numerically and these persist with the MLE. We briefly describe how the equation is integrated 
numerically in Appendix \ref{sec:orig}. We begin by plotting the Lane-Emden function $\theta(\xi)$ for various values of $\Upsilon\sim 10^{-1}$ in 
figure \ref{fig:lesolms}. The deviations from the GR curve are indistinguishable when $\Upsilon\lsim 0.1$. One can see visible changes in the 
shape of the function and the intersect on the $\xi$-axis (i.e. the value of $\xi_R$), namely that solutions with larger $\Upsilon$ have larger 
values of $\xi_R$\footnote{One should be careful to note that this does not necessarily imply that the stars have larger radii since this is 
degenerate with changing the central density and $K$. Indeed, if one wishes to compare two stars then it is important to decide which quantities to 
fix and which to vary in order to make a meaningful comparison.}. Indeed, plotting $\bar{\omega}_R/\omega_R$ and 
$\bar{\xi}_R/\xi_R$ as a function of $\Upsilon$ in figure \ref{fig:om} shows that one expects large signals when $\Upsilon\gsim0.1$. That 
being said, one should not draw too many conclusions directly from $\theta(\xi)$ and $\xi_R$. 



\begin{figure}[ht]\centering
\includegraphics[width=0.45\textwidth]{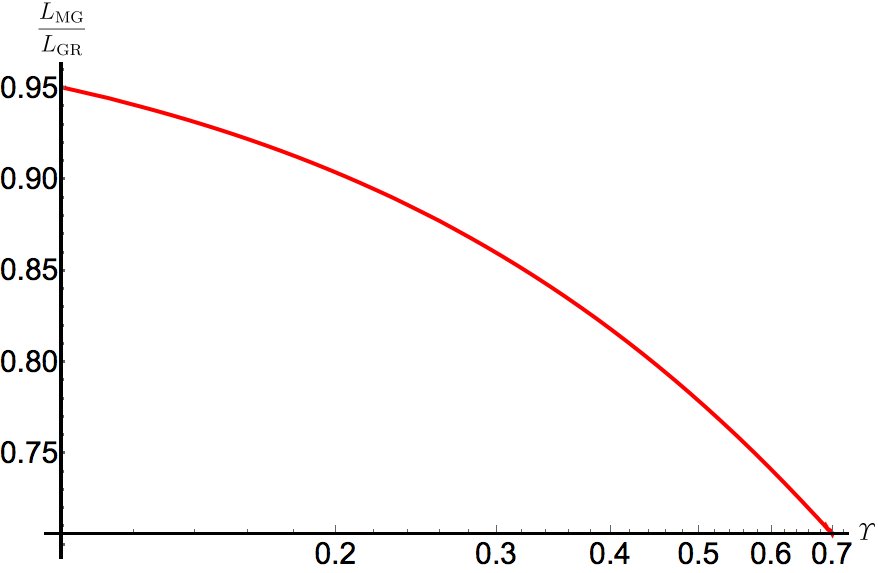}
\caption{The luminosity enhancement for a $1M_\odot$ star as a function of $\Upsilon$.}\label{fig:fixm}
\end{figure}

\begin{figure}[ht]\centering
\includegraphics[width=0.45\textwidth]{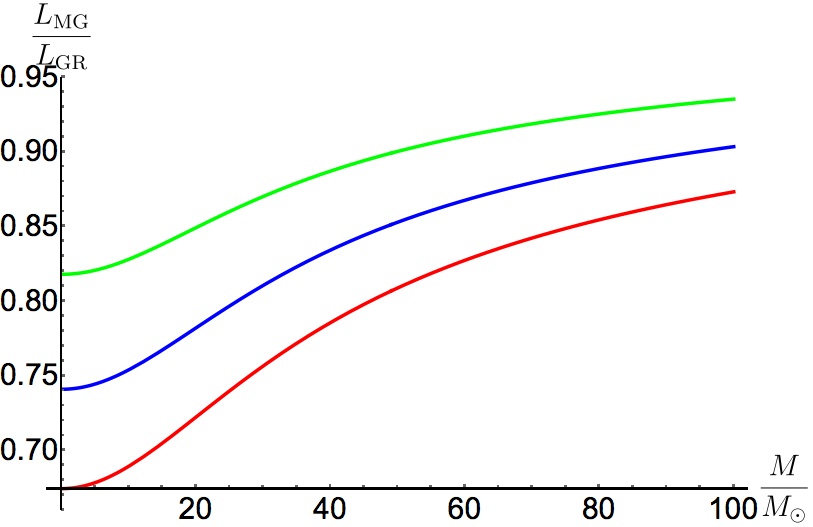}
\caption{The luminosity enhancement as a function of mass for $\Upsilon=0.8$ (red), $\Upsilon=0.6$ (blue) and $\Upsilon=0.4$ (green).}\label{fig:fixu}
\end{figure}

Using these solutions, we can solve equation (\ref{eq:Equart}) to find $\beta(M)$, which can be used in equation (\ref{eq:lum}) to find the ratio of 
the luminosity predicted in the new theory to that predicted by GR at fixed mass. In figure \ref{fig:fixm} we plot this ratio for a $1M_\odot$ star 
as a function of $\Upsilon$. Once can see that there is a visible decrease in this ratio around $\Upsilon\sim 10^{-1}$. In figure \ref{fig:fixu} we 
plot the same ratio as a function of mass for three fixed values of $\Upsilon$. One can see that the modifications affect low mass stars to a 
greater extent than high mass stars. This is because low mass stars are gas pressure-supported whereas high mass stars are radiation 
pressure-supported. The luminosity of gas pressure-supported stars is more sensitive to the strength of gravity than radiation pressure-supported 
stars (see \cite{Davis:2011qf} for a discussion on this), hence the turn towards GR at high masses. The stars in the new theory are dimmer by a 
factor of $10\%$ for $\Upsilon\sim0.2$ and the ratio increases steadily towards $1$ when $\Upsilon$ is decreased. In particular, the ratio is greater 
than $0.95$ when $\Upsilon\lsim0.1$. Very dim stars of 1$M_\odot$  are clearly at odds with observations of the solar 
brightness and so one can tentatively place the constraint $\Upsilon\lsim \mathcal{O}(1)$\footnote{This is only tentative because polytropic 
models are subject to the caveats mentioned above and these effects may be diminished by or be degenerate with non-gravitational physics not included 
in our model. Furthermore, it may possible to achieve the observed solar luminosity in more complicated models by changing unmeasured quantities such 
as the solar equation of state or the neutrino loss rate. Indeed, we will see in the next section that the effect of increasing $\Upsilon$ is 
degenerate with increasing the metallicity.}. Said another way, since the cosmology of this theory has yet to be worked out, any theory with a 
best-fitting model that requires $\Upsilon\gsim \mathcal{O}(1)$ is likely in tension with astrophysical constraints.




\subsection{Realistic Stars}

The simple semi-analytic model above suffers from several drawbacks. Firstly, it does not include any non-gravitational physics. Secondly, it 
predicts that the surface temperature, one of the most important observational properties of stars, is zero\footnote{One can see from equation 
(\ref{eq:beta}) that $T\propto\theta$ and since the star's radius is defined by $\theta(\xi_R)=0$ the temperature is identically zero at the 
stellar radius. More realistic stellar models define the effective temperature as the temperature of the photosphere, which is the radius at which 
the 
optical depth is $2/3$.} and so the modification to the stars colour, if any, is unknown. Finally, it is unable to treat post-main-sequence stars 
such 
as red giants due to their complex composition and the absence of any convective physics in the model.

For these reasons, we have modified the publicly available code MESA \cite{Paxton:2010ji,Paxton:2013pj} to solve the modified stellar structure 
equation (\ref{eq:HSE}). MESA is a self-consistent stellar evolution code that solves the stellar structure equations coupled to convection recipes, 
nuclear reaction networks, atmosphere models and radiative transfer processes. It is able to produce stellar models that are realistic and can be 
compared with observational data. Indeed, it has already been used to place the most stringent constraints on chameleon theories 
\cite{Davis:2011qf,Jain:2012tn,Sakstein:2013pda} to date. Gravitational physics only appears in the MESA code via the hydrostatic equilibrium 
equation. All other equations encode the effects of processes such as convection or nuclear burning, which are not directly controlled by the theory 
of gravity. For this reason, the only change we have made is to replace the GR equation with equation (\ref{eq:HSE}).

\begin{figure}[ht]\centering
\includegraphics[width=0.45\textwidth]{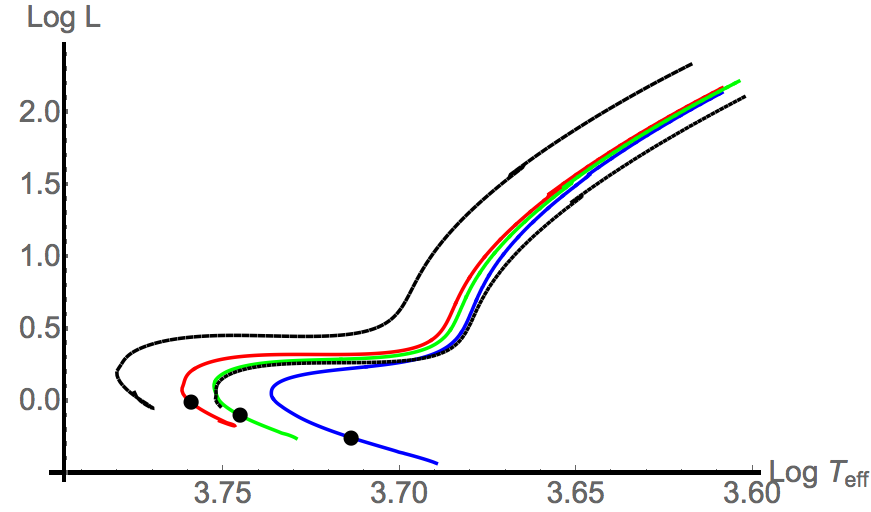}
\caption{The HR diagram for a solar mass star. The red curve is the GR track when the metallicity is solar ($Z=0.02$) and the blue and green tracks 
show the evolution of the same star when $\Upsilon=0.3$ and $\Upsilon=0.1$ respectively. The black dashed (upper) and dotted (lower) lines show the 
tracks for the same star when the theory of gravity is GR and $Z=0.01$ and $0.03$ respectively. The black circles indicate the point on 
the track when the age of the star is $4.6\times10^9$ yr. This is 
not shown for the $Z=0.01$ and $0.03$ stars.}\label{fig:hr1m}
\end{figure}

As an example, in figure \ref{fig:hr1m} we plot the Hertzprung-Russell (HR) diagram for a solar mass star. Included are the tracks for a star 
with solar metallicity ($Z=0.02$) as predicted by GR and $\GT$-galileons with 
various values of $\Upsilon$ indicated in the caption. One can see that the main effect is in the temperature rather than the 
luminosity. Main-sequence stars have a smaller effective temperature than stars at the same point on the HR track when the modification is stronger. 
An examination of the tracks shows that the effect is far more pronounced in main-sequence rather than red giant stars. As with any astrophysical 
system, one must worry about degeneracies. Stars with higher metallicities have smaller effective temperatures\footnote{This is because they have 
more free electrons and the opacity is larger.} at fixed luminosity and this is one potential source of degeneracy. Indeed, figure \ref{fig:hr1m} 
also includes the GR tracks for stars with $Z=0.01$ and $Z=0.03$. One can see that the main-sequence track for a $\GT$ star with $Z=0.02$, 
$\Upsilon=0.1$ is nearly coincident with the main-sequence track for a GR star with $Z=0.03$. On the main-sequence, the effect of $\GT$-galileons is 
to mimic stars of larger metallicity. This is not the case for post-main sequence stars because the tracks converge to a value very close to the GR 
prediction whilst the higher metallicity GR star still has a smaller effective temperature. Finally, we indicate the point on each track when the 
age of the star is solar age. One can see that stars with stronger modifications evolve more slowly and are less ephemeral than their GR 
counterparts (at fixed metallicity).

\begin{figure}[ht]\centering
\includegraphics[width=0.45\textwidth]{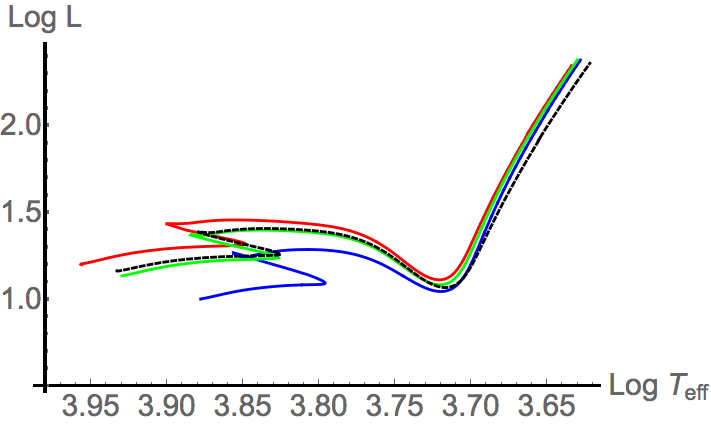}
\caption{The HR diagram for a 2$M_\odot$ star. The red curve is the GR track when the metallicity is solar ($Z=0.02$) and the blue and green tracks 
show the evolution of the same star when $\Upsilon=0.3$ and $\Upsilon=0.1$ respectively. The black dashed line shows the 
track for the same star when the theory of gravity is GR and $Z=0.03$. }\label{fig:hr2m}
\end{figure}

As a final example. we plot the HR tracks for 2$M_\odot$ stars in figure \ref{fig:hr2m}. Shown are the tracks for $Z=0.02$ stars in GR and 
$\GT$-galileons with $\Upsilon=0.3$ and $\Upsilon=0.1$. One can see a similar effect to that found for solar mass stars where the effective 
temperature is smaller with increasing $\Upsilon$ but the decrease in the luminosity is also more pronounced. This is in contrast to what we found in 
the previous section. Higher mass stars have a composition that is far less uniform than low mass stars and are not as well-described by polytropic 
models. If one is simply changing the value of $\GN$ then the change in the luminosity is due entirely to the change in $\beta$, however, since 
equation (\ref{eq:HSE}) depends primarily on the internal structure of the star it is possible that polytropic models miss important contributions 
coming from non-uniform compositions. Finally, we have also plotted the track for a GR star with $Z=0.03$. One can again see that the main-sequence 
portion of this track is nearly coincident with the $\GT$-galileon star of solar metallicity, showing that the degeneracy is still present for higher 
mass stars. Again, the two tracks differ in the post-Main-sequence phase.  

\section{Galaxy Tests}
\label{sec:galtests}
Galaxies are another astrophysical object that can potentially exhibit effects due to their extended dark matter profile. In what follows, we will 
investigate two potential signatures of the breaking of the Vainshtein mechanism: rotation curves and lensing. In all cases we will assume that the 
dark matter is distributed according to the the Navarro-Frenk-White (NFW) profile \cite{Navarro:1995iw}:
\begin{align}
 \rho(r) &= \frac{\rho_{\rm s}}{\frac{r}{r\sss}\left(1+\frac{r}{r\sss}\right)^2}.
\end{align}
We will always examine Milky Way-like galaxies and so we will take $r\sss=20$ kPc and $\rho\sss= 6.68$ 
$M_\odot\textrm{kPc}^{-3}$ in order to provide concrete numbers. These are typical values for the Milky Way (assuming a halo mass of 
$10^{12}M_\odot$) \cite{Neto:2007vq,Xue:2008se}.

\subsection{Rotation Curves}

In general relativity, the rotation curve of a galaxy is a direct probe of the derivative of the gravitational potential and is hence a probe of the 
interior mass distribution. Here, we will take the density profile as given\footnote{One should ideally run new simulations in this model to verify 
that NFW is still a good fit but this is clearly beyond the scope of the present work.} and use the modified equations for $\Phi$ and $\Psi$ to 
predict the new galactic properties given this profile.

\begin{figure}[ht]\centering
\includegraphics[width=0.45\textwidth]{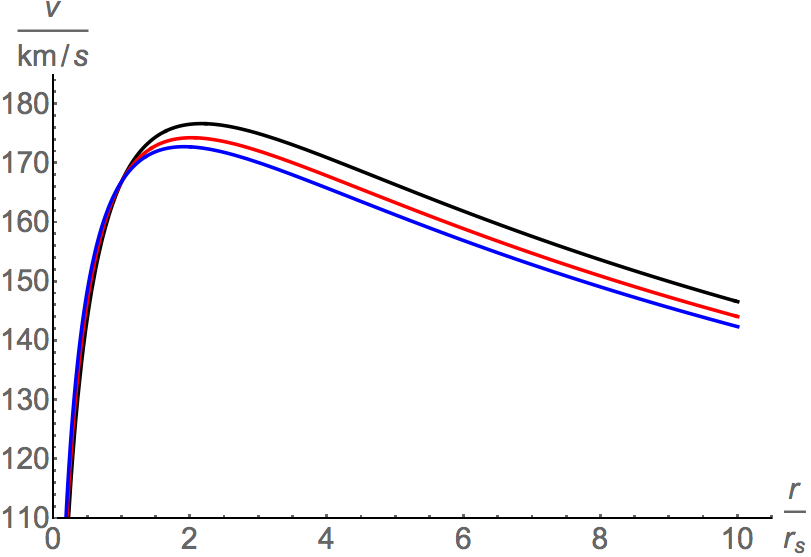}
\caption{The rotation curves for the NFW profile with $r\sss=20$ kPc and $\rho\sss= 6.68$. The GR prediction is shown in black and the red and blue 
curves correspond to $\Upsilon = 0.3$ and $\Upsilon = 0.5$ respectively.}\label{fig:rotNFW}
\end{figure}

Assuming circular motion, the velocity and radius of a test body inside the halo are related via
\begin{equation}
\frac{v^2}{r} =  \frac{\dd \Phi}{\dd r}
\end{equation}
and so using equation (\ref{eq:dPhi}) one finds
\begin{align}
 v^2&=\frac{4\pi \GN 
r\sss^3\rho\sss}{r}\left[\ln\left(1+\frac{r}{r\sss}\right)-\left(1+\frac{r\sss}{r}\right)^{-1}+\frac{\Upsilon}{4}
\frac{\left(\frac { r\sss}{r}-1\right)}{\left(1+\frac{r\sss}{r}\right)^{3}}\right]
\end{align}
The rotation curve for a Milky Way-like galaxy is plotted in figure \ref{fig:rotNFW} for various values of $\Upsilon$ indicated in the caption. One 
can see that the discrepancy between the two theories is minimal at small radii and so the part of the curve measured using stellar velocities is 
largely unchanged. The discrepancy is more pronounced at larger radii where the curves begin to flatten. Here, $\GT$ theories with larger values of 
$\Upsilon$ predict that objects orbit with smaller circular velocities than predicted with GR. The velocities in this portion of the curve (in the 
Milky Way) are typically measured using HI emission lines. 



\subsection{Strong Lensing}

\begin{figure}[ht]\centering
\includegraphics[width=0.45\textwidth]{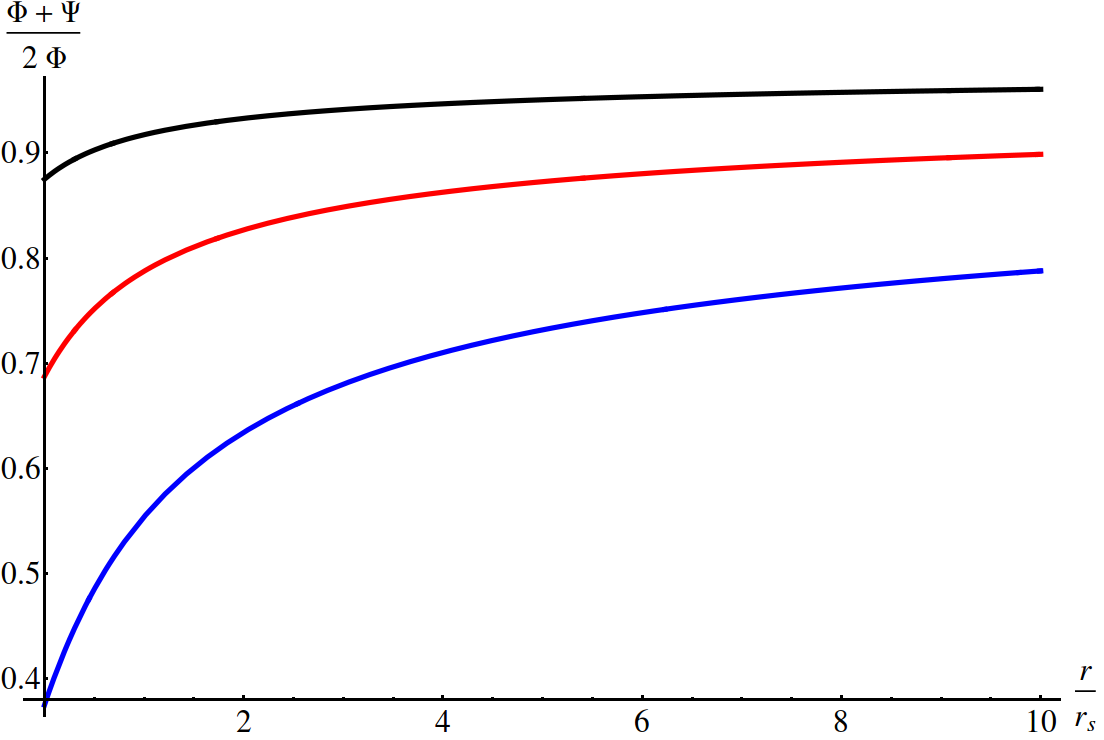}
\caption{The lensing potential predicted by the NFW profile with $r\sss=20$ kPc and $\rho\sss= 6.68$. The GR prediction for this ratio is unity. The 
$\GT$-galileon predictions with $\Upsilon=0.1$, $\Upsilon=0.3$ and $\Upsilon=0.5$ are plotted using black, red and blue curves 
respectively.}\label{fig:lens}
\end{figure}

The bending of light inside a source is governed by the lensing potential $(\Phi+\Psi)$. Using equations (\ref{eq:dPhi}) and (\ref{eq:dPsi}) one 
finds
\begin{align}
 \Phi+\Psi=-\frac{8\pi \GN 
r\sss^3\rho\sss}{r}\left[\ln\left(1+\frac{r}{r\sss}\right)-\frac{\Upsilon}{8}\frac{\left(6+5\frac{r\sss}{r}\right)}{\left(1+\frac{r\sss}{r}
\right)^ { 2 } }\right].
\end{align}
In figure \ref{fig:lens} we plot the ratio $(\Phi+\Psi)/2\Phi$, which quantifies the amount by which light is bent by gravity relative to the 
gravitational force felt by objects moving at non-relativistic velocities, for various values of $\Upsilon$ indicated in the caption. GR predicts 
that $\Psi=\Phi$ so that this ratio is unity. One can see from the figure that this ratio is less than unity in $\GT$-galileon theories and 
decreases with increasing values of $\Upsilon$. This means that the lensing mass is smaller than the dynamical mass and so strong lensing could be 
used as a novel probe. Indeed, \cite{Schwab:2009nz} have already used strong lensing, which probes the weak field metric on kilo-parsec scales, to 
constrain 
deviations of the parameter $\gamma$, defined by
\begin{equation}
\dd s^2=-\left(1+2\frac{\GN M}{r}\right)\dd t^2+\left(1-2\gamma\frac{GM}{r}\right)\dd x^2,
\end{equation}
from unity to the 5\% level. Their constraint cannot be used to constrain our model parameters since $\Phi$ and $\Psi$ are not inverse-square in our 
model. Nonetheless, this serves as an example of how astrophysical measurements such as strong lensing can probe this theory of gravity.

\section{Discussion and Conclusions}\label{sec:concs}

In this paper we have studied the non-relativistic limit of a subset of beyond Horndeski theories, the covariantisation of the quartic galileon or, 
as we have dubbed it, the $\GT$-galileon. We have found that whilst the Vainshtein mechanism is active in suppressing gradients of the scalar degree 
of freedom, the equations of motion for the metric potentials are different from the Poisson equation. In particular, there are extra terms inside 
sources of finite extent proportional to the first and second derivatives of the mass. These terms act to weaken the strength of gravity provided 
that the density decreases outward, as is the case for most astrophysical sources.

This opens up the possibility of testing these theories using the astrophysics of finite sources and in this paper we have taken the first steps 
towards identifying new and novel probes. The strength of the deviations can be quantified by a single number, $\Upsilon$, which is a combination of 
the time-derivative of the cosmological scalar and the new mass scale $\Lambda$ appearing in the quartic galileon Lagrangian. To-date, the specific 
theory has yet to be constrained using other, more conventional probes, and so, a priori, there are no constraints on $\Upsilon$.

One promising probe is the equilibrium structure of stars. Indeed, these have already been used to place the most stringent constraints 
on chameleon-like theories to-date. We have derived the modified hydrostatic equilibrium equation and have used it to study both simple semi-analytic 
polytropic stars and fully realistic stars by updating the stellar structure code MESA. In the former case, the modified Lane-Emden equation was 
derived for a general polytropic index before specialising to the case of main-sequence stars. The weakening of gravity causes stars to be dimmer 
than their GR equivalents at fixed mass. Full MESA simulations revealed that the surface temperature is also lower and so one would generally expect 
stars of fixed mass to be dimmer and redder than their GR counterparts. We also found that the $\GT$-stars evolve at a slower rate than the 
equivalent GR stars. As with any astrophysical test of gravity, there are degeneracies with non-gravitational astrophysical processes. We identified 
a degeneracy with metallicity where a $\GT$-star of fixed mass and metallicity mimics a GR star of the same mass with a larger metallicity on the 
main sequence. This is not true on the red giant branch, where the $\GT$-star tracks are nearly identical to the GR tracks for stars of the same 
metallicity. This then suggests a potential probe of these theories whereby the metallicity (and possibly age) of a globular cluster inferred from 
the main-sequence portion of the HR diagram would differ from that inferred using the red giant branch.

Next, we turned our attention to galactic probes. We calculated the rotation curves for Milky Way-like NFW halos and found that non-relativistic 
objects orbit with smaller circular velocities in $\GT$-galileon theories compared with GR. The effect is minimal on the portion of the curve that 
rises steeply but can be large when the curves flatten. The lensing potentials were also calculated and it was found that the lensing mass is 
smaller than the dynamical mass. The difference can be at the 10\% level or greater for certain parameter choices and so this could potentially be 
probed using strong lensing.

Before concluding, it is worth commenting that the specific model studied here is known to have a gradient instability \cite{Kase:2014yya} at late 
times. Whereas this is not catastrophic for the theory, it is worrying. One can cure this by adding more Horndeski (and possibly beyond Horndeski) 
terms. In terms of our calculations here, this changes very little. Following the analysis of \cite{Kobayashi:2014ida}, no new terms in the equation 
of motion for $\Phi$ are generated and all that changes is the definition of $\Upsilon$ in terms of the fundamental model parameters. Our predictions 
for non-relativistic objects (stars and rotation curves) are therefore robust to this change since we have quoted all of our results in terms of 
$\Upsilon$. The only difference is the introduction of a new parameter in the equation of motion for $\Psi$, by which we mean 
$\Upsilon\rightarrow\tilde{\Upsilon}\ne \Upsilon$ in equation (\ref{eq:dPsi}) whilst $\Upsilon$ remains in (\ref{eq:dPhi}). In this case, the lensing 
potential would depend on two parameters but would look qualitatively the same. This is only the case when the Lagrangian contains quintic terms 
(either Horndeski or beyond Horndeski) and couplings of the scalar to the Ricci scalar proportional to $X$ (see \cite{Kobayashi:2014ida}, Appendix 
2). 

The breaking of the Vainshtein mechanism is only present when the cosmological scalar has a non-vanishing time-derivative. All of the effects 
presented here were non-negligible when $\Upsilon\gsim\mathcal{O}(1)$ and so any theory whose best-fit cosmology predicts values of $\Upsilon$ in this 
range can potentially be probed using astrophysics.

Finally, we conclude by discussing future avenues for testing the breaking of the Vainshtein mechanism. Here we have only studied the equilibrium 
structure of stars but this is only probed using the surface properties. Stellar oscillations and helioseismology allows one to probe the structure 
of 
stars and contains far more information. Indeed, the strongest constraints on $f(R)$ theories comes from stellar pulsations \cite{Jain:2012tn}. Since 
the theory predicts that our own Sun is subject to the breaking of the Vainshtein mechanism, one potential probe would be to look at radial and 
non-radial modes of stellar oscillations. We have also seen that the tracks in the HR diagram for $\GT$-stars mimic those in GR with higher 
metallicities and so one potential signature may arise if larger mass stars cross the instability strip at different locations at fixed metallicity. 
This would manifest as a change in the period-luminosity-mass-colour relation for Cepheid and RR-Lyrae variable stars. Another interesting 
possibility 
is convective stars, which have $n=1.5$. Examining the modified Lane-Emden equation reveals that, for these stars, it contains a term proportional to 
$\theta(\xi)^{-1/2}$, which diverges at the stellar radius. This may suggest that fully convective stars may not exist in these theories. White dwarf 
stars are also subject to the modifications and it is possible to 
extend 
our polytropic analysis\footnote{White dwarf stars are not polytropes except in certain limits but can be treated by deforming the Lane-Emden 
equation.} to derive their new properties and the Chandrasekhar mass to see if it is modified.

All of these effects could potentially probe the entire class of beyond Horndeski models and here we have presented the tools required to do so, 
using the $\GT$-galileon as a specific example.

\section*{Acknowledgements}
We would like to thank Olivier Asin for pointing out a typo in equation \eqref{eq:MLE} which led to figures \ref{fig:lesolms}--\ref{fig:fixu} being incorrect in the published version of this work. We are grateful to Bill Paxton and the MESA community for answering our many questions and to Thomas Collett for several enlightening discussions. We 
would like to thank Alexandre Barreira, Tsutomu Kobayashi, Baojiu Li, Yuki Watanabe and Daisuke Yamauchi for several correspondences. We thank Eugeny 
Babichev, David Langlois and Ryo Saito for useful discussions about the exterior solutions.
\bibliography{ref}

\appendix

\section{Derivation of the Non-relativistic Limit}\label{sec:eoms}
In this Appendix we derive the equations of motion for the scalar field perturbation $\pi$ and the gravitation perturbations $\Phi$ and $\Psi$ in 
the non-relativistic limit following the approach developed in the Horndeski theory \cite{Kimura:2011dc, Koyama:2013paa}. We first ignore terms 
involving time-derivatives of these fields. We then assume that the perturbations are small and neglect non-linear interactions containing 
higher-order powers of these fields and their first-order derivatives. On the other hand we keep all terms with second order derivatives that preserve 
the Galileon symmetry. In addition, unlike the Horndeski theory, the equation of motion for $\pi$ in our theory involves a third order derivative of 
$\Psi$, which breaks the Galileon symmetry. The scalar field equation and the Einstein equations give
\begin{widetext}
\begin{align}
 \nabla^2 \Psi &+ \frac{5 \epsilon}{4}
\Big[(\nabla^2 \pi)^2 - (\nabla_i \nabla_j \pi)(\nabla^i \nabla^j \pi)  \Big]
=4 \pi G \rho, \\
\nabla^2 \Phi - \nabla^2 \Psi&  -\frac{\epsilon}{4} \Big[  
(\nabla^2 \pi)^2 + 3 (\nabla_i \nabla_j \pi)(\nabla^i \nabla^j \pi) + 4 (\nabla_i \nabla^2 \pi)(\nabla^i \pi)
\Big] =0, \\
  \nabla^2 \pi 
&-\frac{2}{\Lambda^4}
\Big[  
(\nabla^2 \pi)^3 - 3 (\nabla^2 \pi)(\nabla_i \nabla_j \pi) (\nabla^i \nabla^j \pi) 
 +2 (\nabla_i \nabla_j \pi)(\nabla_k \nabla^i \pi) 
(\nabla^k \nabla^j \pi) \Big] + \\\nonumber &\!\!\!\!
\varepsilon \Big[ 
5 (\nabla^2 \Phi)(\nabla^2 \pi) 
 - 5 (\nabla_i \nabla_j \Phi)(\nabla^i \nabla^j \pi)+
(\nabla^2 \Psi)(\nabla^2 \pi)
 +(\nabla_i \nabla_j \Psi)(\nabla^i \nabla^j \pi) 
+ 2 (\nabla_i \nabla^2 \Psi)(\nabla^i \pi)
\Big] =0. 
\end{align}
\end{widetext}
Imposing the spherical symmetry and integrating by parts, these equations reduce to equations (\ref{eq:z}), (\ref{eq:y}) and (\ref{eq:x}).

\section{The Vainshtein Mechanism Outside Finite Sources}\label{sec:VM}

In this Appendix we will examine the exterior profile of the metric potentials \footnote{We would like to thank Eugeny Babichev, David Langlois and 
Ryo Saito for useful discussions, which helped us identify the incorrect use of the boundary conditions an earlier verison of this work.}
 
We begin by discussing the boundary conditions. The assumptions of spherical symmetry require us to impose the boundary 
conditions $\pi^\prime(0)=\Phi^\prime(0)=\Psi^\prime(0)=0$. Equations (\ref{eq:x})--(\ref{eq:y}) were found by integrating a set of 
second-order equations once and dividing by $r^3$. In theory, these contain arbitrary integration constants on the right hand side of the form 
$C_i/r^3$, but since these diverge as $r\rightarrow0$ we must have $C_i=0$ and so these terms are absent. The same is not true outside of the source. 
The exterior solution can contain terms that scale as $B_i/r^3$ since the solution is valid for $r\ge R$ only.
The boundary conditions at the stellar radius fix these integration constants. 
Now $R$ is typically defined such that $\rho(R)=0$. In section 
\ref{sec:ss} we found that we had to impose two boundary conditions at the centre of the star\footnote{This is true in general. One requires $\dd 
P/\dd r(0)=0$ due to spherical symmetry and the central pressure must also be specified.} and since the equations are second-order, this leaves no 
freedom to fix any conditions at the stellar surface. In particular, we are not free to set $\rho\ppp(R)=0$ by hand. 
We will use the notation 
$\rho\ppp\equiv\rho\ppp(R)$. 


Accounting for the three integration constants $B_i$, the equations of motion outside the source are 
\begin{align}
 z+\frac{5\varepsilon}{2}x^2-\frac{B_1}{r^3}&=0,\label{eq:zo}\\
 y-z-\frac{\varepsilon}{2}x^2-\varepsilon x\left(rx\ppp+x\right)-\frac{B_2}{r^3}&=0\quad\textrm{and}\label{eq:yo}\\
  x-4\frac{x^3}{\Lambda^4}+10\varepsilon xy+2\varepsilon x\left(2z+rz\ppp\right)-\frac{B_3}{r^3}&=0.\label{eq:xo}
\end{align}
After some 
algebra, one can combine these to find a single equation for $x$:
\begin{equation}\label{eq:xeqo}
  x -4\left(1+5\Upsilon\right)\frac{x^3}{\Lambda^4}+\left(8B_1+10B_2\right)\frac{\sqrt{\Upsilon}x}{\Lambda^2r^3}+\frac{B_3}{r^3}=0.
\end{equation}
This is a cubic equation for $x$, which in general has three distinct solutions. 
 
Since equations (\ref{eq:zo})--(\ref{eq:xo}) do not contain any $\delta$-function sources we expect that $B_3=0$ both inside and outside the source 
to avoid discontinuities in $x$ and $x^\prime$. We therefore make the ansatz that $B_3=0$ and check that this indeed leads to the correct solution 
that satisfies all of the boundary conditions. Thus, the solution is self-consistent. Solving equation (\ref{eq:xeqo}) gives
\begin{equation}\label{eq:xsolphant}
 x^2=\left(\frac{4B_1+5B_2}{2+10\Upsilon}\right) \frac{\Upsilon^{\frac{1}{2}}\Lambda^2}{r^3}+\mathcal{O}(r^{3/2}).
\end{equation}
Equations (\ref{eq:zo}) and (\ref{eq:yo}) contain $x\ppp$ and $z\ppp$ and so we must require that $x$ and $z$ are continuous at the stellar radius. 
This fixes $B_1$ and $B_2$ as  
\begin{equation}
B_1 = G M, \quad B_2=0.
\label{Bi}
\end{equation}
One can then substitute this back into equations (\ref{eq:zo}) and (\ref{eq:yo}) to find expressions for $\Phi\ppp$ and $\Psi\ppp$:
\begin{align}
 \frac{\dd \Phi}{\dd r}&= \frac{GM}{(1+5\Upsilon)r^2}
\quad\textrm{and}\\
 \frac{\dd \Psi}{\dd r}&=\frac{GM}{(1+5\Upsilon)r^2}.
\end{align}
The PPN parameter is defined in (\ref{eq:gammadef}) and one finds
\begin{equation}
 \gamma=1.
\end{equation}
This shows that the definition of Newton's constant (\ref{eq:GNdef}) is indeed correct. 

Finally, we must check for the consistency of the assumption $B_3=0$. Using equations (\ref{eq:xeqo}) and (\ref{Bi}) we find that the solution for 
$x$ is given by 
\begin{equation}
x^2 =\frac{\Lambda^2 (r^3 \Lambda^2 + 8 G M \Upsilon^{1/2})}{4r^3 (1+ 5 \Upsilon)}
\end{equation}
at the stellar radius. This exactly matches the solution of equation (\ref{eq:X}) and therefore confirms that our assumption that $B_3=0$ is 
self-consistent. 

\section{Behaviour of the Modified Lane-Emden equation near the Origin}\label{sec:orig}

In this Appendix we briefly discuss the numerical method of integrating the MLE. The MLE is a second-order ordinary differential equation (ODE) and 
so one may n\"{a}ively expect this to be a simple task whereby an explicit iterative numerical method such as the Runge-Kutta family of methods can 
be used with the two boundary conditions at $\xi=0$ to construct the full solution. Let us begin by discussing the Lane-Emden equation. This is 
singular at the origin and so one typically starts the integration at some small number $\delta\ll1$. This is fine for most ODEs, however the 
Lane-Emden equation possesses a special \textit{homology} symmetry whereby several solutions are related through an appropriate transformation of 
$\xi$ 
and $\theta$ (see \cite{chandrasekhar2012introduction,1987A&A...177..117H,Heinzle:2002sk}). We are not interested in the precise form of these 
transformations here but one of their key implications is that the second boundary condition is redundant and the Lane-Emden equation may be recast 
as 
a first-order ODE. This means that by starting with initial conditions at some small $\delta$ and not $\xi=0$ one may evolve along a completely 
different class of solutions to the physical ones. In order to circumvent this problem, one can expand the Lane-Emden function near the origin as
\begin{equation}\label{eq:thexpo}
 \theta(\xi)=1-\alpha\xi^2+\beta\xi^4+\mathcal{O}\left(\xi^6\right)
\end{equation}
and one finds
\begin{align}
 \alpha &= \frac{1}{6}\quad\textrm{and}\\
 \beta&=\frac{n}{120}.
\end{align}
The equation is then numerically solved by evaluating $\theta(\xi=\delta)$ and $\theta^\prime(\xi=\delta)$ using equation (\ref{eq:thexpo}) with 
these coefficients.

Returning to the MLE, one finds
\begin{align}
 \alpha &= \frac{1}{6}+\frac{\Upsilon}{4}\quad\textrm{and}\\
 \beta&=\frac{n}{240}\left(2+13\Upsilon+15\Upsilon^2\right).
\end{align}
In this work, we use these coefficients with $\delta=10^{-30}$ to solve the MLE numerically.

\end{document}